\documentclass[12pt]{iopart}
\usepackage{float}
\usepackage{times}
\usepackage{graphicx}
\usepackage{subfigure}
\usepackage{color}
\usepackage[usenames,dvipsnames,svgnames,table]{xcolor}
\usepackage[utf8]{inputenc}

\begin{document}

\title[Influence of the growth conditions on the magnetism of SrFe$_{12}$O$_{19}$ thin films...]{Influence of the growth conditions on the magnetism of SrFe$_{12}$O$_{19}$ thin films and the behavior of Co/SrFe$_{12}$O$_{19}$ bilayers}

\author{G.D. Soria$^1$, J. F. Marco$^1$, A. Mandziak$^{1,2}$, S. Sánchez-Cortés$^3$, M. Sánchez-Arenillas$^1$, J. E. Prieto$^1$, J. Dávalos$^1$, M. Foerster$^2$, L. Aballe$^2$, J. López-Sánchez$^4$, J. C. Guzmán-Mínguez$^5$, C. Granados-Miralles$^5$, J. de la Figuera$^1$ and A. Quesada$^5$}

\address{$^1$Instituto de Química Física ``Rocasolano'', CSIC, Madrid E-28006, Spain}
\address{$^2$Alba Synchrotron Light Facility, CELLS, Barcelona E-08290, Spain}
\address{$^3$Instituto de Estructura de la Materia, CSIC, Madrid E-28006, Spain}
\address{$^4$Instituto de Magnetismo Aplicado, UCM, Madrid E-28230, Spain}
\address{$^5$Instituto de Cerámica y Vidrio, CSIC, Madrid E-28049, Spain}
\ead{gdelgadosoria@iqfr.csic.es}

\begin{abstract}
SrFe$_{12}$O$_{19}$ (SFO) films grown on Si (100) substrates by radio-frequency magnetron sputtering have been characterized in terms of composition, structural and magnetic properties by a combination of microscopy, diffraction and spectroscopy techniques. Mössbauer spectroscopy was used to determine the orientation of the films magnetization, which was found to be controlled by both the sputtering power and the thickness of the films. Additionally, the coupling between the SFO films and a deposited cobalt overlayer was studied by means of synchrotron-based spectromicroscopy techniques. A structural coupling at the SFO/Co interface is suggested to account for the expetimental observations. Micromagnetic simulations were performed in order to reproduce the experimental behaviour of the system.
\end{abstract}

\section{Introduction}

Permanent magnets are used in a variety of applications, such as generators and motors in the automotive industry, computer engineering, medical technology and household appliances \cite{coey_permanent_2012}. Many of these magnets contain rare earths \cite{sugimoto_current_2011}, critical elements whose extraction is environmentally harmful and that present a price volatility risk \cite{sprecher_life_2014}. Their replacement by cheaper and more environmentallly friendly materials is therefore sought \cite{chin_permanent_2000,tyrman_structural_2015}. In our case, we have focused on magnetically hard strontium hexaferrite SrFe$_{12}$O$_{19}$ (SFO) as the base for alternative permanent magnets. Structurally, it is composed by spinel and rocksalt blocks \cite{ObradorsJSSC1988,Brabers,PullarProgMatSci2012}. All iron cations are in the Fe$^{3+}$ oxidation state. It has a ferrimagnetic configuration with five different cation environments for iron as described in the Wyckoff notation: three octahedral sites (12k,4f2 and 2a), one tetrahedral site (4f1) and one bipyramidal one (2b) per formula unit. The atomic arrangement of this particular ferrite results in a high magnetocrystalline anisotropy and a high coercive field. However, its saturation magnetization is moderate, M$_{s}$ = 92-74 Am$^{2}$Kg$^{-1}$ \cite{PullarProgMatSci2012}. Strontium hexaferrite in the shape of nanoparticles, platelets, powders and thin films \cite{ghasemi_structural_2008,soria_strontium_2019,hessien_controlling_2008,masoudpanah_synthesis_2012} has been studied in detail to determine its structural and magnetic properties.

It is well-known that the rigid exchange-coupling between a magnetically hard and a soft material can lead to a larger remanent magnetization while avoiding a high coercitivity loss \cite{al-omari_magnetic_1995,skomski_giant_1993,kaushik_exchange_2008,kang_magnetic_2015}. As a consequence, the system energy product (BH)$_{max}$ can be enhanced and several bilayer systems have been previously studied for this purpose:  CoSm/FeCo\cite{al-omari_magnetic_1995}, FePtB\cite{kaushik_exchange_2008}, SrFe$_{12}$O$_{19}$/Fe \cite{kang_magnetic_2015}. In turn, the exchange-coupling at the hard-soft interface may also be non-rigid, as in a spring-magnet, which causes the soft layer to reverse at low fields \cite{quesada_exchange-spring_2018,fullerton_hard/soft_1999,jiang_new_2005}. A third possibility is that both layers are exchange-decoupled and only dipolar interactions mediate.

In this work, we aim at further understanding the nature of the magnetic coupling at hard-soft interfaces involving hexaferrites. In previous unpublished studies \cite{Adrian}, we looked into the magnetic coupling between SFO platelets having out-of-plane magnetization and a cobalt overlayer. We observed no correlation between the magnetic domains of the Co layer and those of the SFO platelets. This lack of registry between both layers was suggested to arise from an absence of exchange-coupling at the interface and from the competition of the shape anisotropy of the metal layer with the magnetodipolar field created by the SFO layer. To avoid the competition with the shape anisotropy of the layer, we devised here an experiment using in-plane magnetized SFO films. Therefore, for this objective, we have deposited a soft cobalt metal layer on top of SrFe$_{12}$O$_{19}$ thin films produced by rf magnetron sputtering and having controlled easy-axis of magnetization within the sample plane.

\section{Experimental Methods}

Strontium hexaferrite films were prepared by radio-frequency (rf) sputtering of a sintered SrFe$_{12}$O$_{19}$ target made from a commercial SFO powder \cite{commercial}. A barium impurity was detected on SFO target by X-ray photoelectron spectra and X-ray absorption spectroscopy. Silicon (100) wafers of 1 mm thickness were chosen as substrates, which were kept at room temperature (RT) during sputtering. The target-to-substrate distance was approximately 60 mm, the base pressure was in the range of 1x10$^{-6}$mbar and 15 min of pre-sputtering were carried out prior to each deposition. A working pressure (Ar/O$_{2}$ with oxygen ratio 2\%) of 7x10$^{-3}$ mbar was used. Under these conditions, series of samples were grown with varying sputtering powers (140 W to 260 W). Post-deposition annealing was performed in air, at 850$^\circ$C during 3 hours \cite{eva,acharya_sputter_1994,acharya_effect_1996}. The determination of the SrFe$_{12}$O$_{19}$ films thickness was done by a Veeco Dektak 150 Profilometer and their surface morphology was analyzed by Atomic Force Microscopy (AFM, Molecular Imaging) in tapping mode. 

X-ray photoelectron spectra (XPS) were recorded with a SPECS Phoibos-150 hemispherical electron energy analyzer under a base pressure of 4x10$^{-9}$ mbar using Al K$\alpha$ radiation. Constant pass energies of 100 and 20 eV were used to record the wide and narrow scan spectra, respectively. The energy scale was referenced to the binding energy (BE) of the C 1s core level of the adventitious contamination layer which was set at 284.6 eV.

Raman spectra were obtained using a micro-Raman Via Renishaw spectrograph, equipped with an electrically cooled CCD camera, and a Leica DM 2500 microscope, under 532 nm laser excitation provided by a Cobolt SambaTM DPSS laser and using a diffraction grating of 1800 l/mm. The laser power reaching the sample was about 2.0 mW and the spectral resolution was 2 cm$^{-1}$. The acquisition time was 20 sec, the range of measurements was 100-1000 cm$^{-1}$ and the objective employed for the measurements was a long distance 50x Leica.

Integral Conversion Electron Mössbauer Spectroscopy (ICEMS) data were acquired at RT using a $^{57}$Co(Rh) source, a parallel plate avalanche counter \cite{gancedo_cems_1991} and a conventional constant acceleration spectrometer. The spectra were computer-fitted and the isomer shift data were referred to the centroid of the spectrum of metallic iron at room temperature.

The crystal structure of the thin films was determined by X-ray powder diffraction (XRD) with a Bruker D8 Advance diffractometer using Cu-K$\alpha$ (1.54 $\AA$) radiation in a $\theta$/$2\theta$ configuration. The measuring step was 0.02$^\circ$/s with a 0.5s measuring time per step.

The magnetic properties were studied with a vibrating sample magnometer attached to a physical property measurement systems (PPSMS Model 6000 controller - Quantum Desing). Hysteresis loops at room temperature were obtained with a maximun applied magnetic field of 5 T both in the plane of the film and perpendicular to the plane configurations.

A cobalt layer (thickness: 2 nm) was grown by molecular beam epitaxy (MBE) on selected SFO films. The resulting bilayer system was analyzed by photoemission electron microscopy (PEEM), X-ray absorption spectroscopy (XAS) and X-ray magnetic circular dichroism (XMCD) at the Co and Fe $L_{2,3}$ edges. These experiments were carried out at the CIRCE beamline\cite{CIRCE}, of the Alba synchrotron.

Micromagnetic simulations were carried out using the mumax code \cite{mumax}. The in-plane size of the simulation cell was 4x4x2 nm$^{3}$ for both layers and several replicas were made to avoid isolated system behavior. The thicknesses of the SFO and Co layers were set as 360 nm and 2 nm, respectively. The exchange stiffness, saturation magnetization and anisotropy values were taken from  the literature \cite{soria_strontium_2019,jiles_introduction_1998,sarau_spin_2007}. The following set of magnetic parameters was thus used as input in the simulations: exchange stiffness of hard and soft phases: A$_{s}$(SFO) = 6x10$^{-12}$Jm$^{-1}$ and A$_{s}$(Co) = 1.5x10$^{-11}$Jm$^{-1}$, respectively; saturation magnetization M$_{s}$(SFO) = 3.8x10$^{5}$Am$^{-1}$ and M$_{s}$(Co) = 1.4x10$^{6}$Am$^{-1}$, respectively; and magnetocrystalline uniaxial anisotropy with magnitude K(SFO) = 3.6x10$^{5}$Jm$^{-3}$ and K(Co) = 4.1x10$^{5}$Jm$^{-3}$, respectively, both having the same easy axis direction (1, 0, 0).

\section{Results and Discussion}

\begin{figure}[htb]
\centerline{\includegraphics[width=0.9\textwidth]{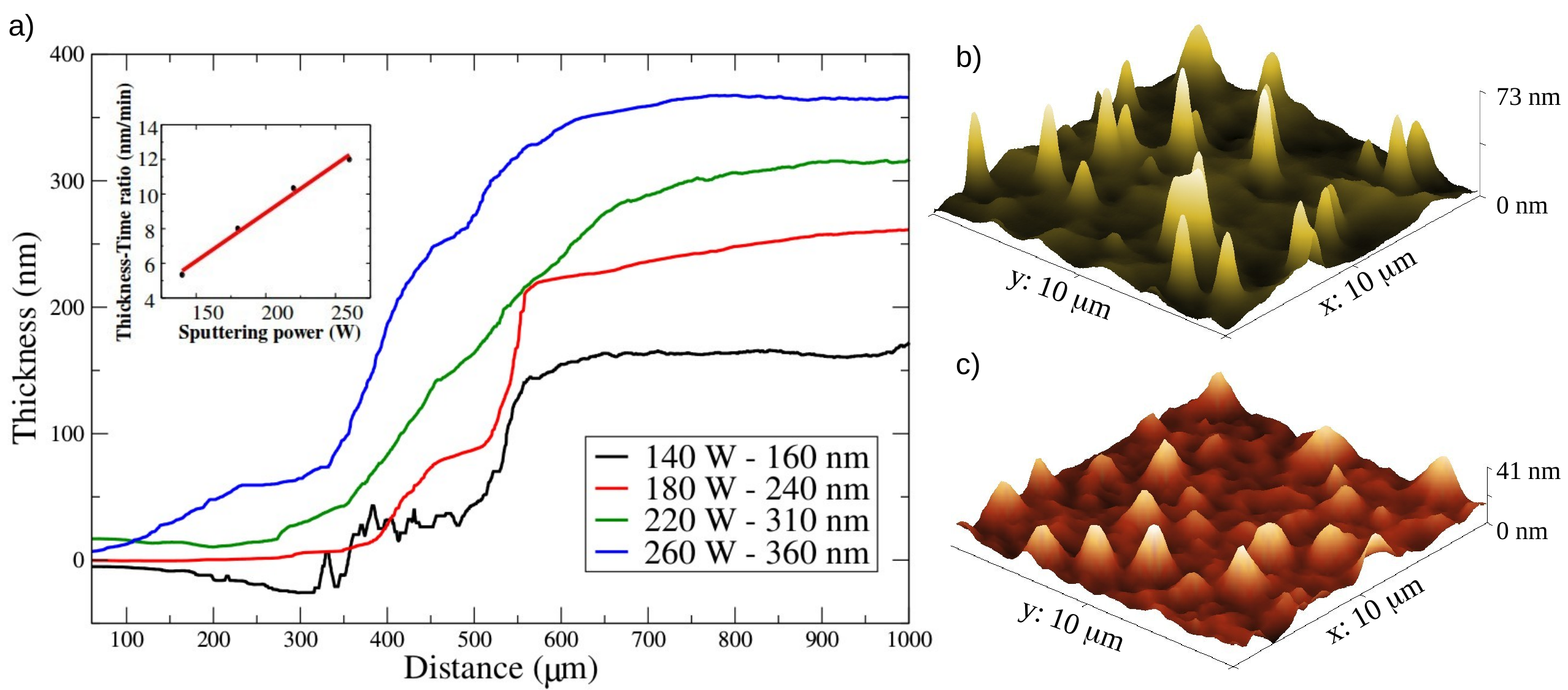}}
\caption{a) SFO films thicknesses measured by a profilometer. Inset: thickness-time ratio to sputtering power (sputtering time, 30 min). Surface of the samples grown at b) 140W and c) 260W.}
\label{AFM}
\end{figure}

Strontium hexaferrite samples were grown by rf magnetron sputtering at different sputtering powers (140 W, 180 W, 220 W and 260 W) for 30 minutes. To determine their thickness, we measured the step between the substrate and the film by means of a profilometer. As can be seen in the inset of figure \ref{AFM}a, the amount of sample deposited in a certain deposition time increases linearly with the sputtering power. 
The surface morphology of the SFO films was studied by AFM. At intermediate powers the growth seems to occur in a columnar way (figure \ref{AFM}b) with tall columns about 73 nm in height separated by hundreds of nanometers. The root mean square roughness (rms) was 8 nm. In contrast, at high powers a flatter denser surface is obtained, although columnar growth is still observed (figure \ref{AFM}c). This sample presents a rms roughness of 12 nm.  According to the results obtained, Cho et al. had already anticipated in his study about barium ferrite thin films that an increase of thickness produces an increase in the surface roughness \cite{cho_thickness_1999}.

\begin{figure}[htb]
\centerline{\includegraphics[width=0.9\textwidth]{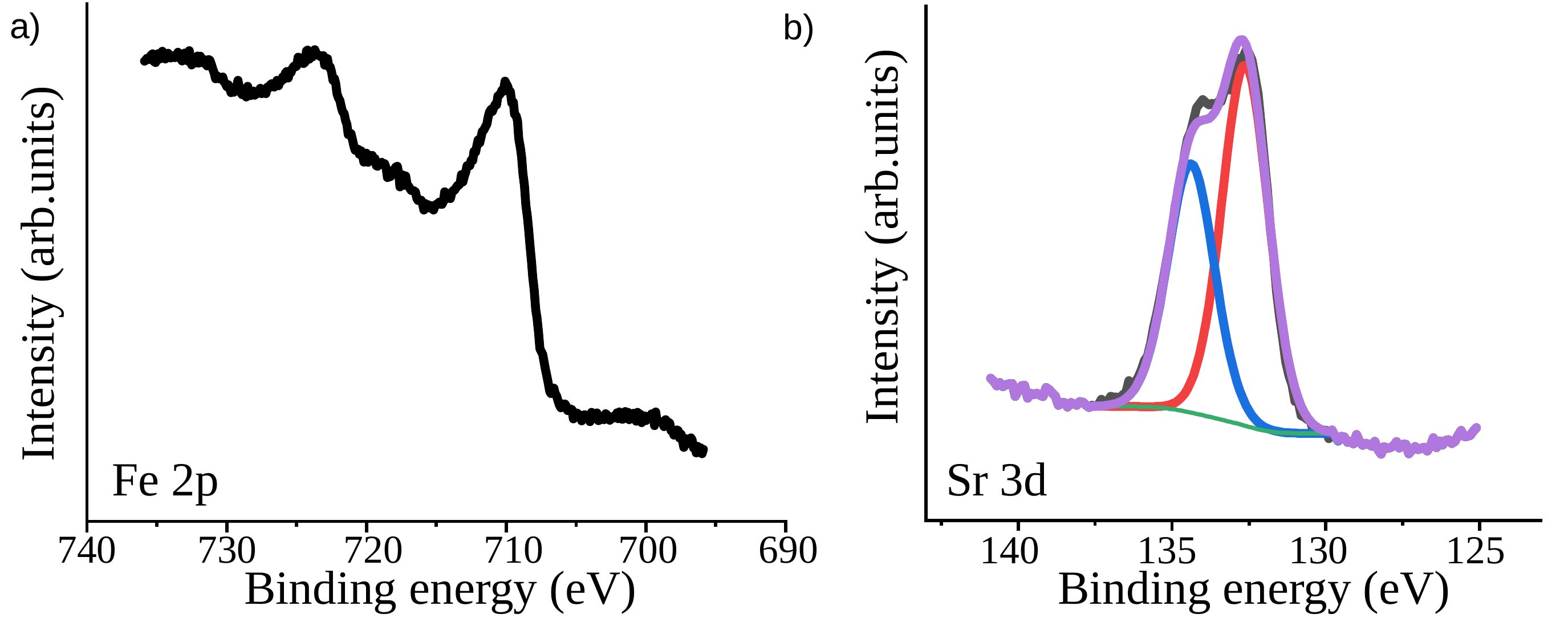}}
\caption{a) Fe 2p and b) Sr 3d XPS spectra of a SFO film deposited at 260 W.}
\label{XPS}
\end{figure}

The elemental composition and the oxidation state of the surface cations were determined by XPS. The Fe 2p spectra recorded from the films consist of a main spin orbit-doublet with binding energies (BE) of the Fe 2p$_{3/2}$ and 2p$_{1/2}$ core levels at 710.2 eV and 723.3 eV, respectively (figure \ref{XPS}a). These binding energies and the presence of a small shake-up satellite at 718.1 eV are all characteristic of the presence of Fe$^{3+}$ \cite{yamashita_analysis_2008,biesinger_resolving_2011}. Figure \ref{XPS}b shows the Sr 3d spectrum from the same sample confirming the presence of Sr$^{2+}$ (BE Sr 3d$_{5/2}$ = 132.6 eV and Sr 3d$_{3/2}$ = 134.4 eV). The data correspond to a thin film deposited at 260 W, but all films gave very similar spectra.

\begin{figure}[htb]
\centerline{\includegraphics[width=0.75\textwidth]{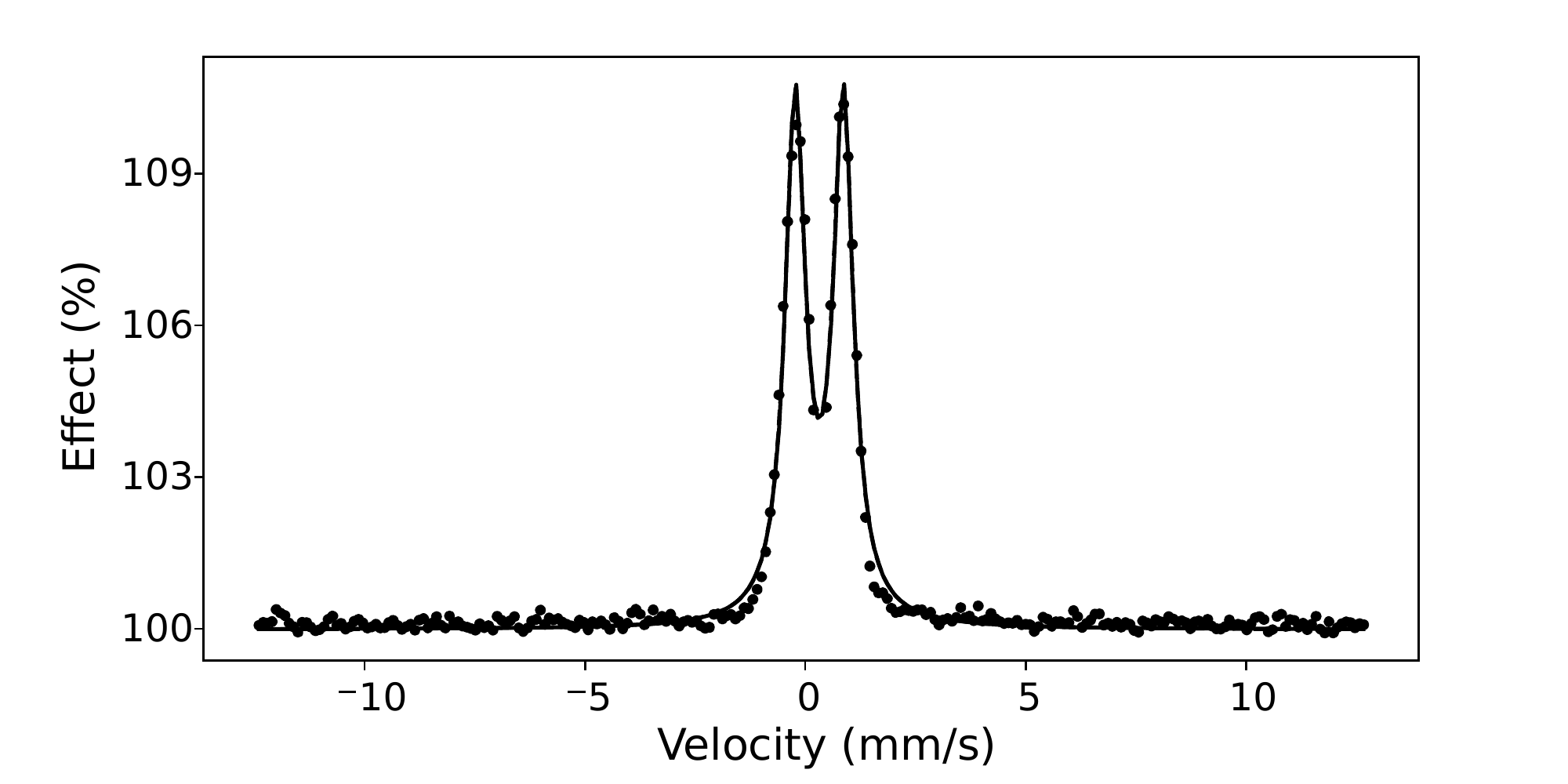}}
\caption{Mössbauer spectrum of a thin film grown by magnetron sputtering at a power of 140 W prior to the annealing treatment.}
\label{Mossbauerdoblete}
\end{figure}

Mössbauer spectroscopy is a very useful technique that allows identifying iron compounds by determining their hyperfine parameters. From a Mössbauer spectrum information about the chemical, structural and magnetic state of a particular iron cation can be obtained. In Figure \ref{Mossbauerdoblete}, the ICEMS spectrum of a freshly deposited SFO film (before annealing) is presented. A paramagnetic doublet is observed instead of the usual five magnetic components expected for SFO\cite{soria_strontium_2019}. The collapse of the magnetic interactions at room temperature is an indication of the poorly crystalline/superparamagnetic character of the as-deposited film and probably arises from the small size of the magnetic grains. The isomer shift ($\delta$ = 0.32 mm/s) is characteristic of Fe$^{3+}$ \cite{menil_systematic_1985}. Since previous works have demonstrated the need for annealing the sample either \textit{in-situ} or \textit{ex-situ} in order to crystallize the SFO phase \cite{morisako_effect_1997,ramamurthy_acharya_preparation_1993, hui_effect_2014}, the films were annealed at 850$^\circ$C for three hours. 

\begin{figure}
\centerline{\includegraphics[width=0.8\textwidth]{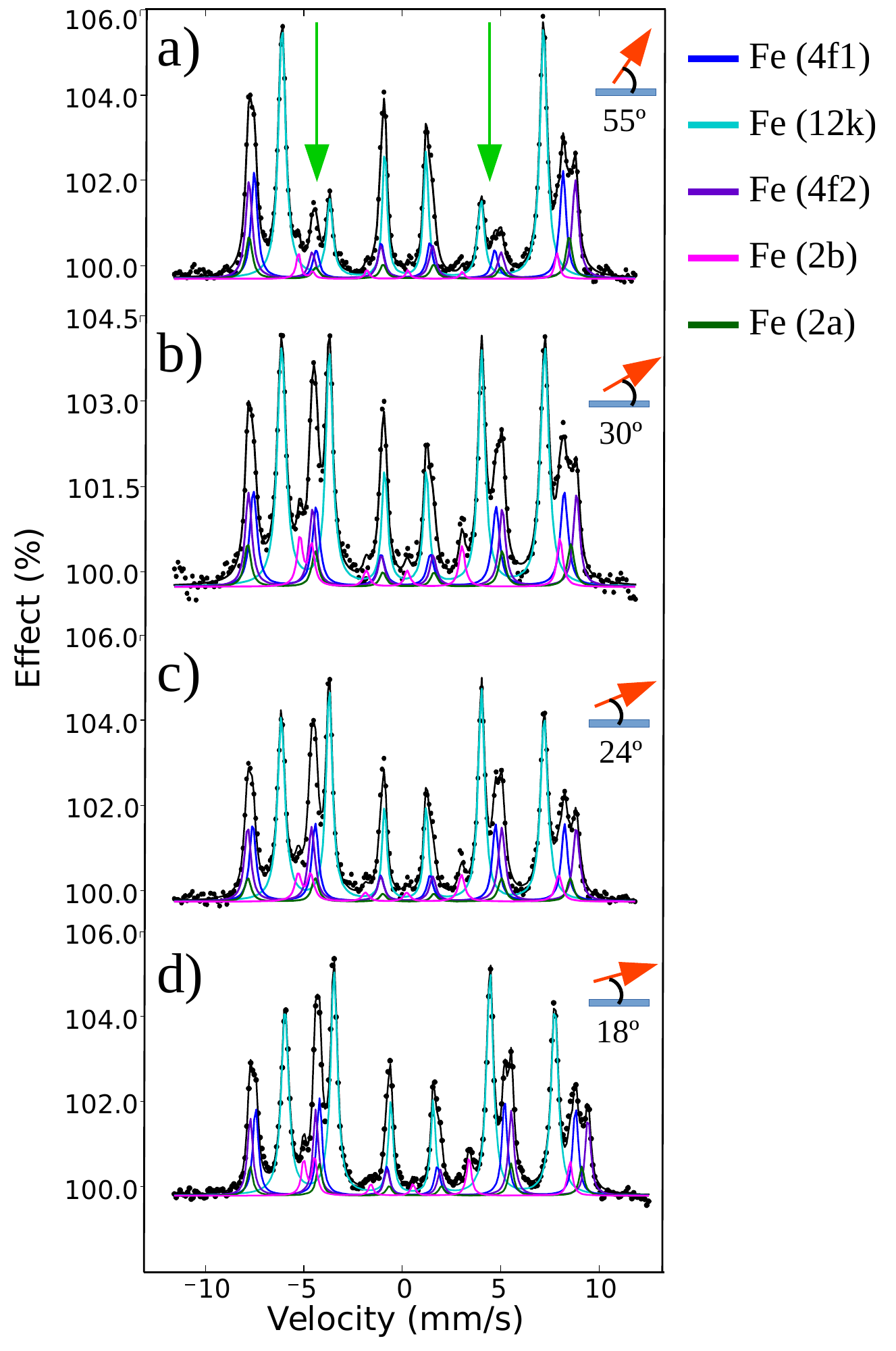}}
\caption{Room temperature ICEMS spectra recorded from SFO films deposited at different magnetron powers: a)140 W, b) 180 W, c) 220 W and d) 260 W. The five sextets in each spectrum correspond to the sites occupied by the iron cations in the SFO structure \cite{soria_strontium_2019}. The green arrows point to the 2 and 5 lines of the sextets that undergo a change in intensity as the sputtering power increases. The arrows on the upper right side of each spectrum symbolize the average angle of magnetization direction in each sample. Note that the angle depicted in the figure is that formed between the hyperfine magnetic field and the surface plane, i.e., the complementary angle of that mentioned in the text.}
\label{Mossbauer}
\end{figure}

Figure \ref{Mossbauer} shows the spectra recorded from the films deposited at different sputtering powers for 3 hours (a) 140 W, b) 180 W, c) 220 W and d) 260 W) followed by the annealing at 850$^\circ$C. All the spectra were fitted to five sextets, each sextet corresponding to Fe$^{3+}$ in a different chemical environment (three octahedral sites -12k, 4f2, 2a-, one bipyramidal site -2b- and a tetrahedral site -4f1-)\cite{ObradorsJSSC1988}. 
The values of the different hyperfine parameters are virtually the same for all films. Table \ref{tableM} collects a representative set of hyperfine parameters which corresponds to the film grown at 260 W (Fig.\ref{Mossbauer}d). The parameters are all characteristic of strontium hexaferrite \cite{BerryJMSL2001,soria_strontium_2019}.

\begin{table}[htb]
\centering
\begin{tabular}{|c|c|c|c|c|c|}
\hline
Site & $\delta$  & 2$\varepsilon$ & H & $\Gamma$ & Area \\
& ($\pm$ 0.03 mms$^{-1}$) & ($\pm$ 0.05 mms$^{-1}$) & ($\pm$ 0.05 T) & ($\pm$ 0.05 mms$^{-1}$) &($\%$) \\
\hline
12k & 0.34 & 0.38 & 41.3 & 0.28 & 53 \\
4f1 & 0.24 & 0.18 & 49.0 & 0.32 & 18 \\
4f2 & 0.37 & 0.28 & 51.7 & 0.32 & 17 \\
2a & 0.33 & 0.04 & 50.7 & 0.32 & 6 \\
2b & 0.27 & 2.22 & 40.8 & 0.28 & 6 \\
\hline
\end{tabular}
\caption{${^{57}}$Fe Mössbauer parameters obtained from the fit of the spectra shown in Figure \ref{Mossbauer}d. The symbols $\delta$ , 2$\varepsilon$, H, $\Gamma$ correspond to isomer shift, quadrupole shift, hyperfine magnetic field and linewidth, respectively.}
\label{tableM}
\end{table}

An important aspect of the ICEMS spectra is the change in the relative intensity of the second and fifth lines of the sextets as a function of the magnetron deposition power. It is well-known that the relative areas of Mössbauer absorption lines reveal the average direction of the magnetization in the sample respect to the $\gamma$ rays direction, which is perpendicular to the sample plane. The spectral areas follow the ratio 3:x:1:1:x:3 where x is a function of $\theta$ (the angle between the direction of the incoming $\gamma$-rays and the direction of the magnetic hyperfine field), x=4sin$^{2}$$\theta$/1+cos$^{2}$$\theta$\cite{greenwood_mossbauer_1971,voogt-assisted_1999}. Therefore, a value x = 0 corresponds to a perpendicular orientation of the magnetization respect to the sample plane ($\theta$ = 0$^\circ$) while a value x = 4 indicates an in-plane orientation ($\theta$ = 90$^\circ$). Inspection of figure \ref{Mossbauer}shows that a power of 140 W (figure \ref{Mossbauer}a), gives x = 0.8, that corresponds to an average angle of 35$^\circ$ and therefore implies a large amount of grains with perpendicular orientation of the magnetization respect to the sample plane. Increasing the sputtering power results in an increase in x, amounting up to x = 3.3 for 260 W (figure \ref{Mossbauer}d) and this indicates an almost in-plane magnetization (average 72$^\circ$). These results clearly demonstrate that the orientation of the magnetization can be tuned by the appropriate choice of the magnetron deposition power, as had been anticipated long ago by Acharya et al \cite{acharya_sputter_1994,acharya_effect_1996,ramamurthy_acharya_magnetic_1997}. 

\begin{figure}[htb]
\centerline{\includegraphics[width=1.0\textwidth]{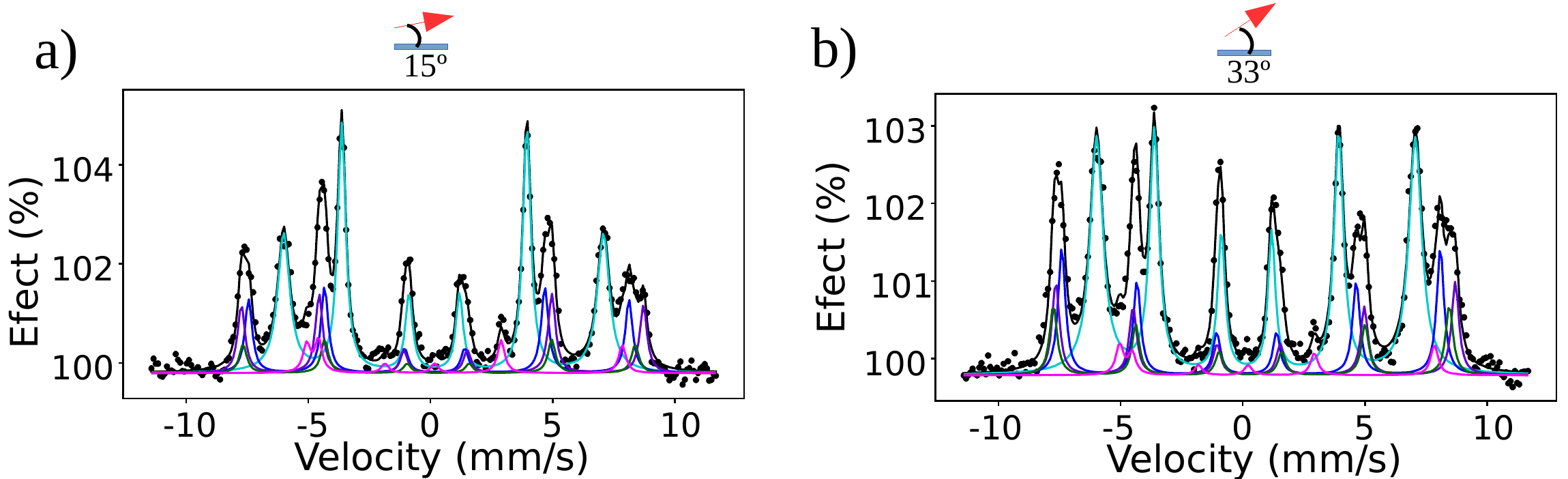}}
\caption{Mössbauer spectra for SFO films: a) Sample a and b) Sample b.}
\label{Supmossbauer}
\end{figure}

In the case of the samples reported here, we have observed that, for a given deposition time, the thickness of the films increases with magnetron sputtering power as expected (figure \ref{AFM}a). Concomitantly, we have observed that the increase in sputtering power/thickness of the film tends to align the magnetization parallel to the sample plane. The work by Cho et al \cite{cho_thickness_1999} had already pointed out the dependence of the magnetization with the film thickness for barium ferrite, BFO (which is isostructural with SFO) on sapphire (001). In that study, they found that at thicknesses less than 100 nm, the BFO grains grew perpendicular to the substrate surface with the c-axis out-of-plane, while at larger thicknesses elongated grains grew at the top of the columnar grains parallel to the surface substrate, their c-axis lying also parallel to the surface plane. Since, as in SFO, the magnetocrystalline anisotropy in barium ferrite leads to a magnetization easy axis along the c-direction, the existence of elongated grains arranged along the sample plane results in a decrease of the magnetization normal to the film. Many studies of Sui's group confirm that the morphology and orientation of the grains affects the magnetic properties of thin films \cite{sui_magnetic_1993,sui_microstructural_1993,xiaoyu_sui_magnetic_1993}. Contrarily to this, Ajan et al \cite{ajan_conversion_1998} found that the orientation of the magnetization in SFO films was not dependent on the film thickness but on the sputtering power and the O$_{2}$/Ar ratio used during deposition. So, SFO films having all the same thickness (240 nm) showed different magnetization orientations. Given these contradictory results, and in order to disentangle the role of the sputtering power and the thickness of the films in the magnetization orientation (in the previous experiments both parameters vary), we prepared samples of thicknesses corresponding to the films whose ICEMS spectra are depicted in Fig.\ref{Mossbauer}a and \ref{Mossbauer}d but produced at different sputtering powers. So, a first sample was deposited at 140 W with a thickness of 360 nm (Sample a) and a second (Sample b) was grown at 260 W with a thickness of 160 nm. The rest of the parameters of the sputtering process were not changed. Subsequently, both samples were also subjected to an annealing treatment at 850$^\circ$C for 3 hours. Figures \ref{Supmossbauer}a and \ref{Supmossbauer}b show the Mössbauer spectra recorder from Sample a and Sample b respectively, which are both characteristic of strontium hexaferrite \cite{BerryJMSL2001,soria_strontium_2019}. The area ratio of the spectral lines obtained from the fit of the spectrum recorded from Sample a was 3:3.5:1:1:3.5:3. This implies that the sample shows a magnetization practically in the plane (magnetization/surface angle of 15 $^\circ$). This is remarkable because if we compare this result with that shown in figure \ref{Mossbauer}a, which corresponds to a film produced with the same sputtering power but having a much smaller thickness, we observe a significant difference in the averaged orientation of the magnetization. In the case of Sample b (Figure \ref{Supmossbauer}b) we obtained an area ratio of 3:2.2:1:1:2.2:3 which corresponds to a magnetization having significant out-of-plane components (average 33$^\circ$). So, despite the low sputtering power used, a thicker film makes the magnetization to be practically in-plane (Sample a) while a larger sputtering power but a smaller thickness (Sample b) gives place to a situation intermediate between those of the films giving place to the spectra depicted in figures \ref{Mossbauer}a and \ref{Mossbauer}b, implying that the thicker film contains significant magnetization components out-of-plane. Therefore, the present results indicate that the sputtering power used during deposition is not, by itself, the main factor that determines the magnetization orientation and that the thickness and morphology of the sample are also crucial. The orientation of the magnetization in thin films depends on the interplay among the shape anisotropy, the interface anisotropies and the magnetocrystalline anisotropy. At this respect, it is interesting to recall the AFM images presented in Fig \ref{AFM}b and \ref{AFM}c. As explained above, thinner films show the presence of isolated, tall columns separated by hundreds of nanometers emerging from a continuous film while thicker films show much shorter columns arranged in a much denser configuration (a similar behaviour has been found in Samples a and b, not shown). It has been reported than in the case of columnar structures having a very low areal density, the shape anisotropy of the columns tends to align their magnetization perpendicular to the plane \cite{yafet_directional_1986}. Although this kind of behaviour could play here some role in the case of the thinner films, we must take into account, however, that strontium hexaferrite shows a very high magnetic anisotropy constant (3.6x10$^{5}$Jm$^{-3}$) and that it is well established that it presents uniaxial anisotropy with the easy magnetization axis along the c axis of the crystalline structure \cite{PullarProgMatSci2012}. We must also consider that our XRD results have shown that the thinner films tend to grow with the c-axis perpendicular to the sample plane while thicker films show a trend to grow with the c-axis parallel to the surface plane. Taken together, the present results indicate that the orientation of the magnetization in these films results from the balance between the shape anisotropy of the columnar grains, the magnetocrystalline anisotropy of the SFO films and the shape anisotropy of the overall deposited film, the magnetocrystalline term being most probably the dominant one. In any case, the results show that using a lower sputtering power the orientation of the magnetization can be tuned by an appropriate choice of the film thickness in a more defined way than using much higher sputtering powers. Work is under way aimed at an in-depth understanding of the growth mechanisms resulting from the sputtering processes as well as to understand the influence of the subsequent annealing treatment exercises on the morphology, the crystallinity and the magnetic properties of the deposited films.
 
\begin{figure}[htb]
\centerline{\includegraphics[width=0.75\textwidth]{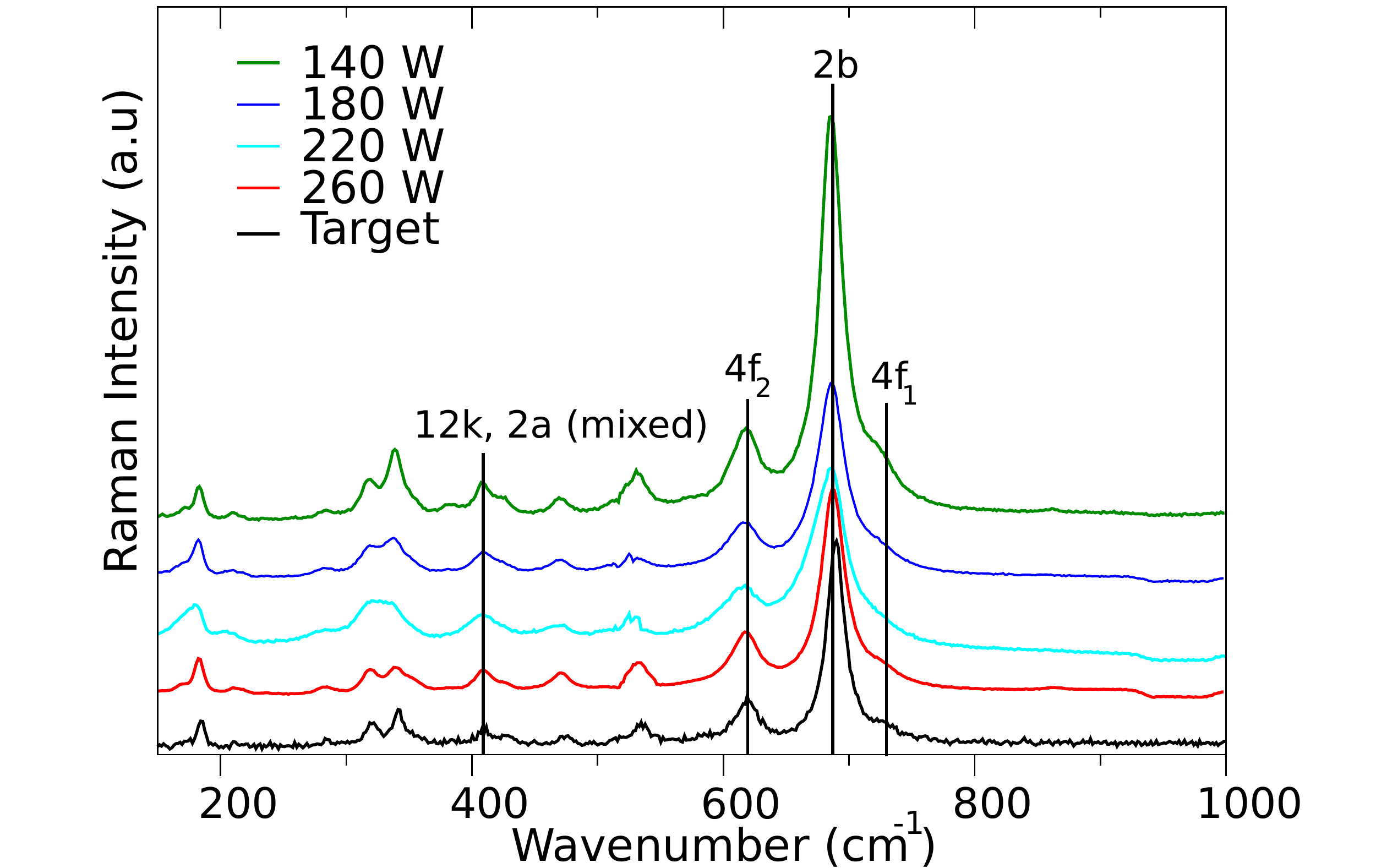}}
\caption{Room-temperature Raman spectra of SrFe$_{12}$O$_{19}$ recorded from the deposited films and commercial sputter target.}
\label{Raman}
\end{figure}

The structural characterization of the films was further carried out by Raman spectroscopy (figure \ref{Raman}). The spectra include bands at 410, 620, 688 and 734 cm$^{-1}$ which are characteristic of strontium hexaferrite \cite{morel_sublattice_2002,kreisel_raman_1999,nethala_investigations_2018,kreisel_raman_1998}. These Raman modes are assigned to vibration modes arising from different chemical enviroments of Fe$^{3+}$. Specifically, the peak observed at 410 cm$^{-1}$ is assigned to the octahedral sites (12k dominated and 2a), the peak at 620 cm$^{-1}$ corresponds to the octahedral sites (4f2) and the peaks at 688 and 734 cm$^{-1}$ are due to the bipyramidal sites (2b) and tetrahedral sites (4f1) sublattice vibration modes, respectively. The Raman spectrum of the commercial strontium ferrite target confirms that the grown films have indeed the composition of SFO as shown also by Mössbauer spectroscopy.

\begin{figure}[htb]
\centerline{\includegraphics[width=0.8\textwidth]{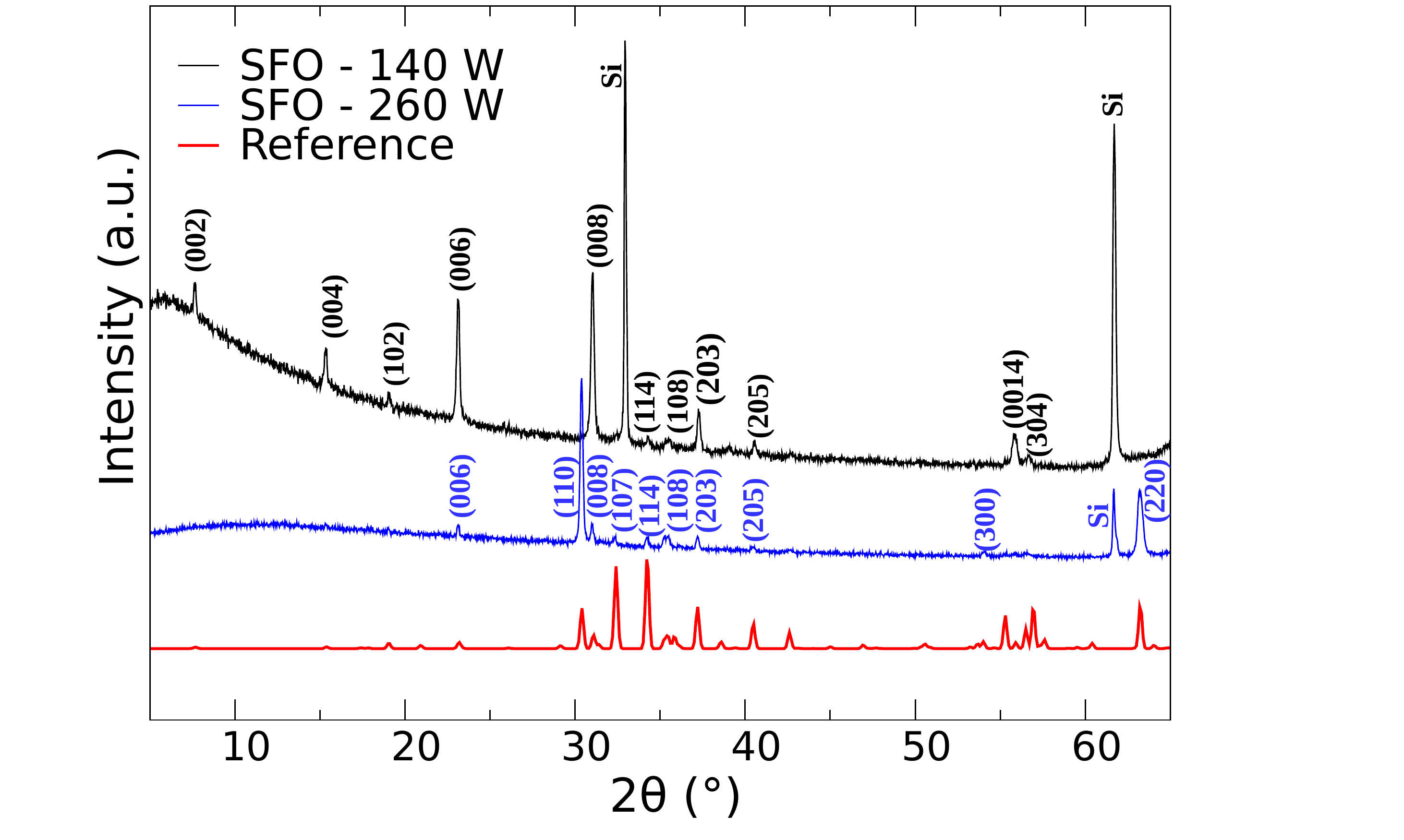}}
\caption{XRD diffraction patterns from the SFO films grown at 140 W (black) and 260 W (blue), together with the simulated pattern (bottom one).}
\label{XRD}
\end{figure}

The crystalline quality and composition of the SFO films were also investigated by XRD. Figure \ref{XRD} presents the XRD patterns recorded from the samples grown at sputtering powers of 140 W and 260 W together with the diffraction pattern of a reference SrFe$_{12}$O$_{19}$ (using 66403 ICSD file \cite{muller_new_1992}). All the diffraction peaks of the SFO films can be indexed according to the reference hexagonal structure of SFO. As seen in the figure, the film deposited at the lowest power shows prominent (00\textit{l}) peaks along [(002), (004), (006),(008) and (0014)] with a few other peaks of less intensity, suggesting a preferential c-axis orientation normal to the film. However, for larger sputtering powers (260W), the (110) and (220) reflections are considerably more intense than those expected for a randomly oriented sample, indicating that this film grows along the 110 direction, i.e., has its c-axis in-plane. 

Therefore, the results obtained by XRD point out that depending of the sputtering power and thickness, the films grow with different preferential crystalline orientations. As we have disscussed before, SFO has its magnetic easy axis along the c axis, so the film magnetization orientation observed in Mössbauer data is consistent with the orientation of the c-axis determined by XRD.

\begin{figure}[htb]
\centerline{\includegraphics[width=0.8\textwidth]{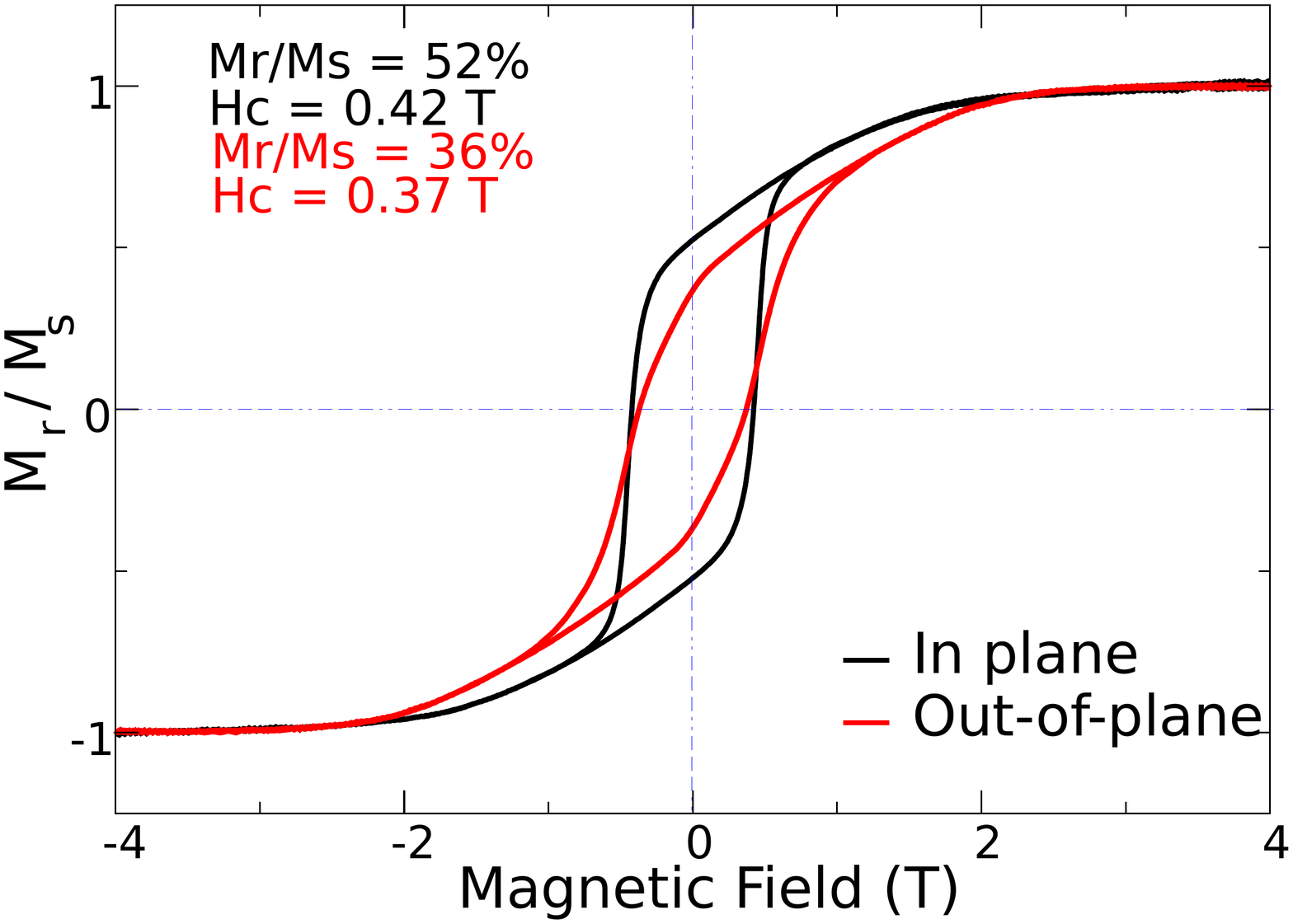}}
\caption{Room-temperature hysteresis loops recorded from SFO film deposited at 260 W after annealing. The black curve was recorded with a magnetic field applied within the plane. Red curve recorded with a magnetic field applied perpendicular to the sample.}
\label{VSM}
\end{figure}

Knowing that the SFO film deposited at 260 W presents magnetization in the plane and, hence, it is adecuate for studying the interaction with a soft layer without the competition of shape anisotropy, its magnetic properties were studied by recording magnetization curves. Figure \ref{VSM} shows RT hysteresis loops. The black curve was measured by applying a magnetic field parallel to the sample plane while the red curve was recorded by applying a magnetic field perpendicular to the sample plane. In the case of the black curve, the measured coercive field is 0.42 T, the saturation magnetization being achieved for an applied field of 1.8 T. From the red curve a smaller coercive field (0.37 T) is measured while the saturation magnetization is reached at a slightly larger applied field (2.3 T). Figure \ref{VSM} also shows a larger remanence magnetization in the black curve than in the red one. These observations confirm that the film has its magnetic easy axis mainly in the sample plane.

Once the suitable sample of strontium hexaferrite was characterized, we proceeded to deposit a cobalt layer of 2 nm thick by molecular beam epitaxy for the study of the coupling between magnetically hard and soft materials in a bilayer system.

\begin{figure}[htb]
\centerline{\includegraphics[width=1\textwidth]{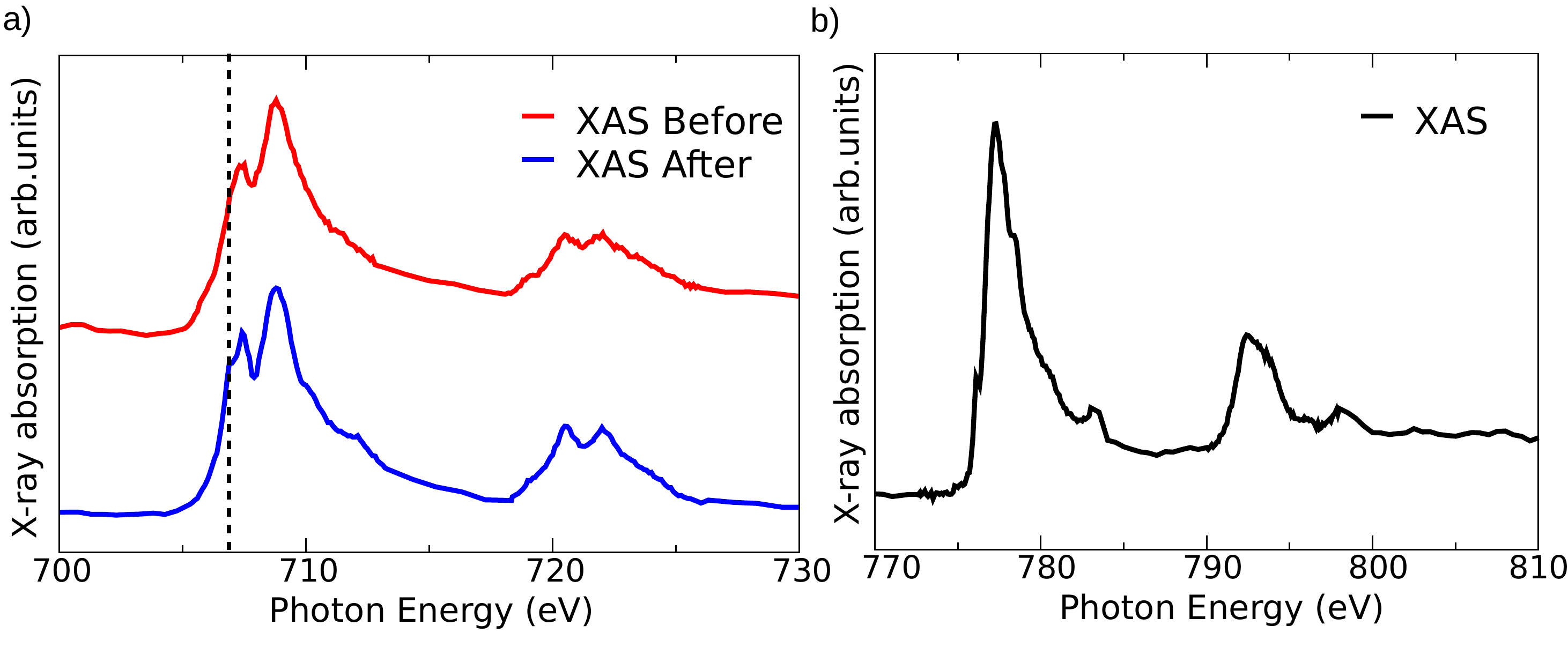}}
\caption{a) XAS spectra at the Fe L$_{2,3}$ edge before and after cobalt exposition. b) XAS spectrum at the Co L$_{2,3}$ edge.}
\label{XAS}
\end{figure}

Figure \ref{XAS}a shows Fe L$_{2,3}$ edge XAS spectra recorded from SFO thin films before and after the deposition of a cobalt overlayer. The spectra are slightly different. An additional feature is observed at a lower photon energy (706.9~eV) in the spectra of the Co-covered SFO films. It can be attributed to the existence of a small Fe$^{2+}$ fraction probably arising from the reduction of some surface Fe$^{3+}$ ions at the SFO/Co interface (\textit{vide infra}). 
The spectrum acquired at the Co L$_{2,3}$ edges~(figure \ref{XAS}b) is quite similar to that shown by metallic cobalt \cite{welke_xmcd_2015}. The spectrum shows, however, some minor peaks at 776.0 eV and 778.5 eV, which are compatible with the presence of some Co$^{2+}$ arising from the oxidation of a fraction of metallic cobalt during deposition \cite{espectrofalso}. It seems that during deposition some interface cobalt atoms react with oxygen atoms of the SFO film and this results on the one hand, on their oxidation to Co$^{2+}$  and, on the other hand, in a concomitant reduction of some Fe$^{3+}$ of the SFO to Fe$^{2+}$. Note also the two peaks appearing beyond the Co L$_{3}$ (784.0 eV) and L$_{2}$ edges (797.5 eV). They correspond to a residual Ba contamination \cite{welke_xmcd_2015,kim_soft_2016,hayakawa_charge-state_2019} stemming from the SFO target used in the magnetron sputtering system. \\

\begin{figure}[htb]
\centerline{\includegraphics[width=1.0\textwidth]{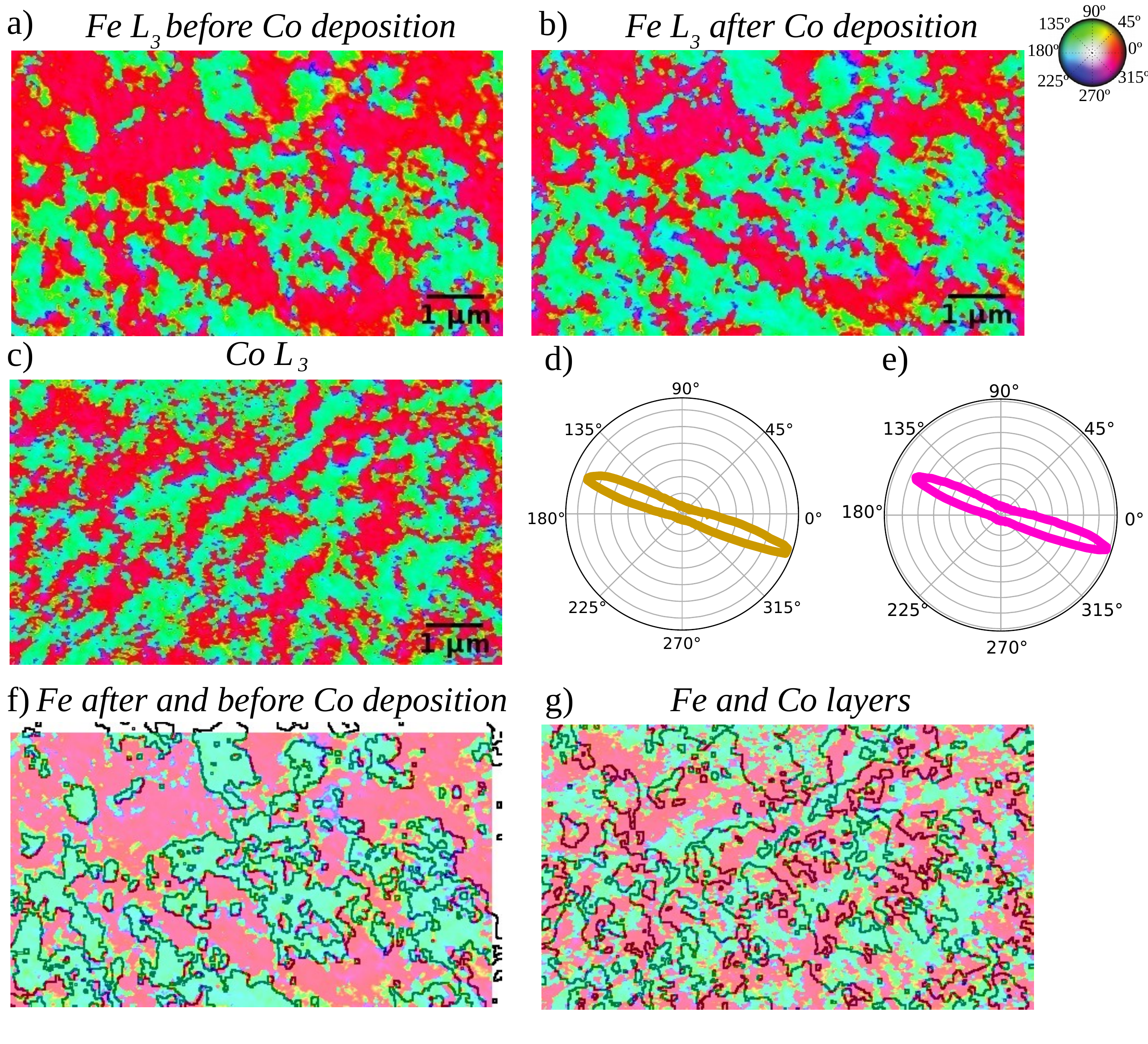}}
\caption{Vector magnetization for: Fe L$_{3}$ edge before a), after b) Co deposition and c) Co L$_{3}$ edge. d) and e) polar plots representing the magnetization distribution in the surface plane extracted from a) and c) images, respectively. f) Overlay images at Fe L$_{3}$ edge before and after cobalt deposition. g) Overlay images at Fe L$_{3}$ edge and Co L$_{3}$ edge after cobalt deposition. Note that the color palette in the upper corner represents the spin direction in the magnetic domains.}
\label{XMCD}
\end{figure}

In order to understand the nature of the coupling between the two magnetic layers (soft cobalt and hard SFO), a 3D magnetization map was measured at the Fe L$_{3}$ edge before cobalt deposition (figure \ref{XMCD}a) and at the Fe L$_{3}$ (figure \ref{XMCD}b) and Co L$_{3}$ edges (figure \ref{XMCD}c) after Co deposition. Since the 3D magnetization vector is proportional to the magnetization in each layer, three XMCD images acquired at different azimuthal angles (0$^\circ$, 60$^\circ$ and 120$^\circ$) were recorded to extract the magnetization distribution at the region of interest on the sample \cite{Sandra}. The iron magnetic domains, which correspond to the ones in the hexaferrite layer, are the same before and after Co deposition (figure \ref{XMCD}a and \ref{XMCD}b). Figure \ref{XMCD}f shows a comparison between the Fe L$_{3}$ edge of SFO overlapping figure \ref{XMCD}b (color patterns) on figure \ref{XMCD}a (domain contours). The most prominent result is that the SFO film shows a uniaxial anisotropy large enough as to show a magnetization vector preferentially aligned in one direction within the film plane (160$^\circ$ and 340$^\circ$), figure \ref{XMCD}d.

The cobalt overlayer domains are unrelated to the iron ones (figure \ref{XMCD}c). Thus, the magnetic domains of the hexaferrite layer are not imposed on the cobalt layer. This result strongly suggests a lack of exchange-coupling between the hard and soft layers. However, it is observed that the easy axis of the cobalt layer is the same as the in-plane easy axis of the hexaferrite layer (figure \ref{XMCD}e), with domains pointing either parallel or antiparallel to the underlying hard domains.

The question then is, given the absence of exchange-coupling, what is the mechanism responsible for the alignment of the easy axes while keeping the domain structures uncorrelated. First, let us consider the hexaferrite structure in the [110] direction, which is the film growth direction. Strontium hexaferrite is a ferrimagnet, with iron cations in some crystallographic sites pointing along the net magnetization direction, and others pointing in the opposite direction. Along (110), there are planes within the unit cell with different populations of each iron cation. Thus the net surface magnetization is opposite along different planes. If some grains of the hexaferrite film present such different terminations, a magnetic coupling with the cobalt layer should give rise to grains where the magnetization of the cobalt, coupled to the net surface magnetization of the hexaferrite, has opposite directions between different grains. However, we believe this scenario is unlikely for several reasons. On the one hand, the film needs to present a substantial number of grains with different terminations, although the crystal termination is often determined by the lowest surface energy. On the other hand, this situation would require a correlation between the domain walls in the cobalt overlayer and the hexaferrite layer, even if the magnetization were coupled either ferromagnetically or antiferromagnetically (depending on the underlaying termination). No such correlation is observed in figure \ref{XMCD}g.

The most likely explanation is that both directions (but not sense) of the magnetization in the two layers are coupled structurally. This could happen in two ways. On one side, the strain imposed at the interface by the ferrite layer could favor a specific growth direction of the cobalt layer that leads to the alignment of the easy axes. On the other, the epitaxial relationship alone could explain this alignment as well. Given the very low thickness of the cobalt layer, our data, especially XRD, are not able to confirm the epitaxial relationship. Thus, we suggest that the coupling between the two layers is structural instead of magnetic. This result is not entirely surprising, as for instance growth of cobalt on the W(110) always produces some uniaxial anisotropy of the cobalt layer \cite{bansmann_surface_2001,pinkvos_spin-polarized_1992} and this does not require different surface terminations.

A consequence of this interpretation, already suggested by the lack of correlation between domain patterns in the two layers, is that in the absence of exchange-coupling, dipolar interactions alone do not lead to the alignment of the spins of the soft layer with the magnetization of the hard one.

\begin{figure}[htb]
\centerline{\includegraphics[width=1.3\textwidth]{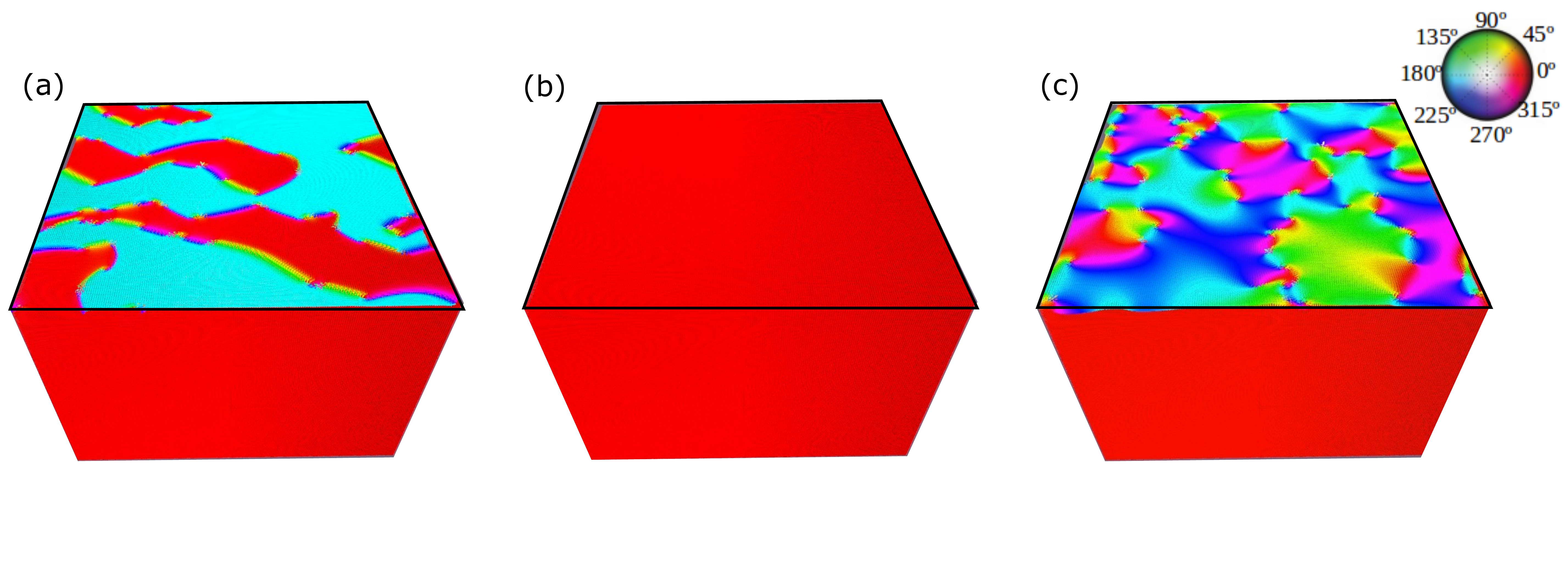}}
\caption{Micromagnetic simulations of bilayer SFO/Co: a) without exchange coupling, b) with 25 $\%$ exchange coupling and c) Co without magnetocrystalline anisotropy and without exchange coupling. Note that the color palette in the upper corner represents the spin direction in the magnetic domains.}
\label{Micromagnetic}
\end{figure}

In order to understand the spin behavior observed in the PEEM images for the bilayer, we have performed micromagnetic simulations \cite{mumax} in a simplified system. Such system consists of a strontium hexaferrite slab having a well-defined in-plane magnetization easy axis (100) covered by a layer of cobalt on top. The cobalt overlayer has its easy axis of magnetization oriented along the same direction (100) as the easy axis of the hexaferrite layers. The thickness of the two layers was set to that shown by the experimental sample studied in PEEM (SFO thin film 360 nm thick with a 2 nm Co overlayer). As initial configuration we used a random multi-domain structure for the cobalt layer and a single domain configuration for the hexaferrite layer. Then, the initial configuration was relaxed for evolving the magnetization as closely as possible to the minimum energy state for the different cases. The first simulation was carried out considering no exchange coupling between the two layers. We observed that the magnetic domains in the soft and the hard magnetic layers are indeed not correlated (figure \ref{Micromagnetic}a)). The hard magnetic layer (SFO) shows only one magnetic domain (in red in the figure) that corresponds to orientation of the spin at 0$^\circ$, while in the soft magnetic layer (Co) there are two magnetic domains (blue and red) that represent the spin in the same directions but in opposite sense, 180$^\circ$ and 0$^\circ$, respectively. The second study was performed incorporating 25 $\%$ interlayer coupling. In this case, the domains in the cobalt and ferrite layers are totally aligned (figure \ref{Micromagnetic}b). For both layers, we can see a unique magnetic domain (red color) which means that all the spins point in the same direction and sense. For the third case, no magnetocrystalline anisotropy in the cobalt layer was set the simulation running in the absence of interchange coupling with the SFO layer. The micromagnetic simulations show that the soft cobalt layer presents magnetic domains within the plane in all cases irrespectively of the orientation of the magnetic domain of the hard ferrite layer (figure \ref{Micromagnetic}c), if there is no exchange coupling between them. There, the strontium hexaferrite layer has a single magnetic domain (red color) while the cobalt layer has magnetic domains with different magnetization directions (rainbow). The simulations thus support that the dipolar coupling is unable to impose the magnetization domain pattern of the hard layer onto the soft layer in the thickness range investigated.

\section{Conclusions}

We have grown and characterized strontium hexaferrite films made by rf magnetron sputtering using different spectroscopy, diffraction and microscopy techniques. Mössbauer spectroscopy allowed the determination of the magnetization orientation of the films. The XRD data have shown that the films are textured and that their structural orientation changes with the sputtering power and thickness. Taken together, the Mössbauer and XRD data suggest that the magnetization of the SFO films is oriented along the c-axis direction, i.e., is mainly defined by the magnetocrystaline anisotropy. After the deposition of a Co overlayer on the SFO films, the XAS spectra at the Co and Fe $L_{2,3}$ edges revealed the presence of a a small amount of Fe$^{2+}$ and Co$^{2+}$ at the SFO/Co interface due to the interaction of the deposited cobalt with oxygen atoms from the SFO surface. Vector magnetization maps show the same uniaxial easy axis for both the SFO film and Co layer. However, the magnetic domains in the cobalt overlayer are not correlated with the magnetic domains in SFO surface. We suggest that the layers are not exchange-coupled, but, there exists a structural coupling between them. This is further supported by comparison with micromagnetic simulations, which confirm that dipolar interactions alone do not lead to an alignment of the soft spins with the hard layer magnetization.

\section*{Acknowledgments}

This work is supported by the Spanish Ministry of Science, Innovation and Universities through Projects RTI2018-095303-B-C51, RTI2018-095303-B-C53 and RTI2018-095303-A-C52 (MCIU/AIE/FEDER,EU) and by the European Comission through Project H2020 No. 720853 (Amphibian) and by the Regional Government of Madrid through project S2018-NMT-4321. C.G-M. acknowledges financial support from MICINN through the "Juan de la Cierva" program (FJC2018-035532-I). Technical support of the technical staff of CIRCE beamline of the ALBA Synchrotron Light Facility is gratefully acknowledged.

\section*{References}
\bibliographystyle{iopart-num}
\bibliography{Bilayer}

\end{document}